\documentclass[10pt,a4paper,twoside]{article}    
\usepackage[T2A]{fontenc}
\usepackage[cp1251]{inputenc}
\usepackage[english,russian]{babel}
\usepackage{amssymb,amsmath,amsfonts}
\usepackage{stfi}
\usepackage{graphicx,epstopdf}
\newcommand{\issueYear}{March 2019}
\newcommand{\issueVolume}{Volume 61, Issue 11}

\begin{document}


\UDK{530.12:531.51:519.711.3}
\PACS{...}
  
\Title%
  {Dynamics of cosmological models with nonlinear classical phantom scalar fields. II. Qualitative analysis and numerical modeling}%
  {}

\Abstract%
  {A detailed qualitative analysis and numerical modeling of the evolution of cosmological models based on nonlinear classical and phantom scalar fields with self-action are performed.  Complete phase portraits of the corresponding dynamical systems and their projections onto the Poincar? sphere are constructed.  It is shown that the phase trajectories of the corresponding dynamical systems can, depending on the parameters of the model of the scalar field, split into bifurcation trajectories along 2, 4, or 6 different dynamical flows.  In the phase space of such systems, regions can appear which are inaccessible for motion. Here phase trajectories of the phantom scalar field wind around one of the symmetric foci (centers) while the phase trajectories of the classical scalar field can have a limit cycle determined by the zero effective energy corresponding to a Euclidean Universe.}%
  {}

\Key%
  {cosmological model, qualitative analysis, phantom scalar field}%
  {}

\Datereceive{July 2, 2018.} 

\Author%
  {Yu.\,G.~Ignat'ev}
  {\textbf{Ignat'ev Yurii Gennadievich}, Doctor of Physics and Mathematics, Professor, Lobachevsky Institute of Mathematics and Mechanics, Kazan Federal University, ul. Kremlyovskaya, 35, Kazan, 420008, Russia.}
  {ignatev-yurii@mail.ru}  
  {}
  {}  

\Author%
  {A.\,A.~Agathonov}
  {\textbf{Agathonov Alexander Alexeevich}, Candidate of Physics and Mathematics, Assistant Professor, Lobachevsky Institute of Mathematics and Mechanics, Kazan Federal University, ul. Kremlyovskaya, 35, Kazan, 420008, Russia.}  
  {a.a.agathonov@gmail.com}
  {}
  {}

\begin{otherlanguage}{english}
\Header


In a preceding paper [1] we formulated a mathematical model of the cosmological evolution of a classical scalar field and a phantom scalar field with self-action. The model reduced to an autonomous system of nonlinear ordinary differential equations in phase space, where it was shown that, depending on the parameters of the field model, the singly-connectedness of the phase plane can be violated.  In the given paper we perform a detailed analysis of the dynamical system formulated in [1].    

\section{SINGULAR POINTS OF A DYNAMICAL SYSTEM}

The singular points of the dynamical system are determined by the equations (see, for example, [2, 3]) 
\begin{equation}
M:\;\;P(x,y) = 0;\;Q(x,y) = 0.
\end{equation}
Since, according to [1, Eqs. (36) and (37)], it is always true at the singular points of the dynamical system that $Z \equiv y = 0$, we obtain the equation 
\begin{equation}
x\left( {{e_2} - {\alpha _m}{x^2}} \right) = 0
\end{equation}
for finding solutions to system of equations (1).  Note that at the singular points of the dynamical system, according to [1, Eq. (18)], 
\begin{equation}H. = 0;
\end{equation}
therefore, all singular points of the dynamical system lie on the inflation trajectory $\Omega  = 1$. Further, for any values of $\alpha_m$ and $\lambda_m \ge 0$, system of algebraic equations (1) always has a trivial solution  
\begin{equation}x = 0;\,\,\,\,y = 0\; \Rightarrow {M_0}(0,0),
\end{equation}
and for  $e_2\alpha > 0$ it has two additional symmetric solutions: 
\begin{equation}x = {x_ \pm } =  \pm \frac{1}{{\sqrt {{e_2}{\alpha _m}} }},\,\,y = 0\quad  \Rightarrow {M_ \pm }\left( {{x_ \pm },0} \right).
\end{equation}
Substituting solutions (4) and (5) into the condition given by Eq. (38) in [1], we obtain a necessary condition for realness of the solutions at the singular points specified by formulas (4) and (5): 
\begin{equation}\left( 4 \right) \to {\lambda _m} \geqslant 0,\,\,(5) \to {\lambda _m} + \frac{1}{{2{\alpha _m}}} \geqslant 0.
\end{equation}

\subsection{Characteristic equation and qualitative analysis in the case near a zero singular point}

We now calculate the derivatives of the functions $P(x,y)$ and $Q(x,y)$ [1, Eqs. (36)] at the zero singular point ${M_0}$ for ${\lambda _m} \geqslant 0$: 
\begin{equation*}
\begin{array}{ll}
  {\displaystyle{{\left. {\frac{{\partial P}}{{\partial x}}} \right|}_{{M_0}}} = 0,}&{\displaystyle{{\left. {\frac{{\partial P}}{{\partial y}}} \right|}_{{M_0}}} = 1,} \\ 
  {\displaystyle{{\left. {\frac{{\partial Q}}{{\partial x}}} \right|}_{{M_0}}} =  - {e_1}{e_2},}&{\displaystyle{{\left. {\frac{{\partial Q}}{{\partial y}}} \right|}_{{M_0}}} =  - \sqrt {3{\lambda _m}} .} 
\end{array}
\end{equation*}

The matrix of the dynamical system at the zero point is     
\begin{equation*}
{A_{{M_0}}} = \left( {\begin{array}{*{20}{c}} 0&1 \\ 
  { - {e_1}{e_2}}&{0 - \sqrt {3{\lambda _m}} } 
\end{array}} \right),
\end{equation*}
and its determinant is 
\begin{equation}   
{\Delta _0} = \det {\left( A \right)_{M0}} = {e_1}{e_2}.
\end{equation}

Thus we obtain the characteristic equation and its roots ${k_ \pm }$: 
\begin{equation} 
\left| {\begin{array}{cc}
  { - k}&1 \\ 
  { - {e_1}{e_2}}&{ - k - \sqrt {3{\lambda _m}} } 
\end{array}} \right| = 0 \Rightarrow {k_ \pm } =  - \frac{{\sqrt {3{\lambda _m}} }}{2} \pm \frac{{\sqrt {3{\lambda _m} - 4{e_1}{e_2}} }}{2}.
\end{equation}

As is well known, the product of the eigenvalues of a matrix is equal to its determinant: 
\begin{equation} 
{k_1}{k_2} = \det \left( A \right).
\end{equation}

Therefore, the signs of the eigenvalues at the zero point, as a consequence of Eq. (7), are completely determined by the index e2.  It is also easy to see that the eigenvalues at the zero singular point are generally independent of the self-action constant, and also of the character of the field (classical / phantom), i.e., of the value of the parameter e1.  Let us consider all possible cases.

\subsubsection{Classical field with attraction ($e_1 = e_2 = 1$) and phantom field with repulsion ($e_1 = e_2 = –1$)}

These two cases are identical at the zero singular point $M_0$.  Here four different situations are possible: 

1a)  The case of zero value of the cosmological constant  
\begin{equation} 
\lambda  = 0 \to k =  \pm i.
\end{equation}

Since the eigenvalues in this case turn out to be purely imaginary, the only singular point [1, Eq. (47)] of the dynamical system [1, Eq. (33)] is the center [2].  In this case, as $\tau  \to  + \infty $ the phase trajectory of the dynamical system winds around this center, completing an infinite number of turns.  

1b) The case of a small value of the cosmological constant  
\begin{equation} 
0 < {\lambda _m} < \frac{4}{3}.
\end{equation}

Here we have two complex-conjugate eigenvalues, where 
\begin{equation} 
\operatorname{Re} \left( k \right) =  - \frac{{\sqrt {3{\lambda _m}} }}{2} < 0.
\end{equation}

In this case, according to the qualitative theory of differential equations, the point $M_0$ (Eqs. (4)) is an attractive focus – in the limit $\tau  \to  + \infty$ all phase trajectories of the dynamical system are twisting spirals, winding around the singular point and in the process completing an infinite number of turns.  This case, in fact, coincides qualitatively with the previous one (Fig. 1).

\begin{figure}[ht!]
\centering
\begin{minipage}[t]{.48\textwidth}
  \centering
  \includegraphics[width=\textwidth]{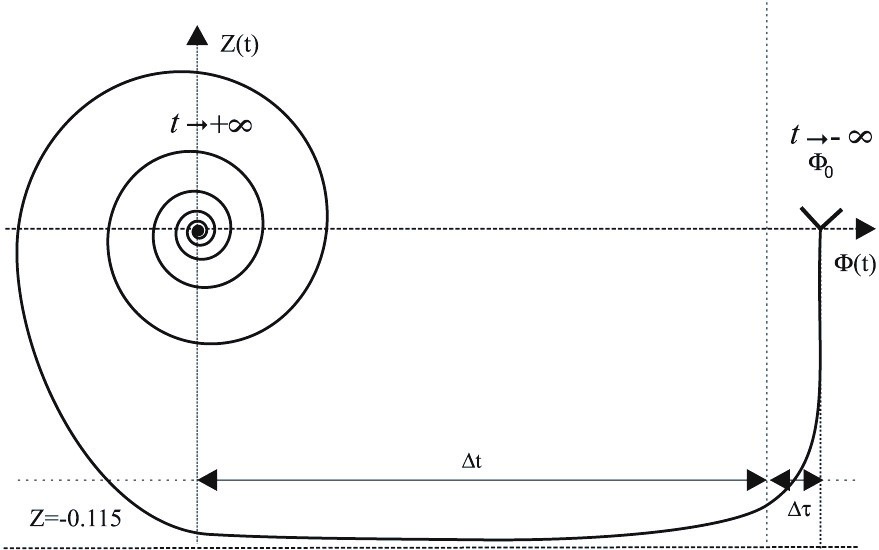}
  \caption{Qualitative form of the phase trajectory of the dynamical system [1, Eq. (33)] for a classical scalar field for ${\lambda _m} < 4/3$: $\Delta \tau$ is the characteristic falloff time of the rate of change of the potential to the bottom of the graph,  ${Z_0} \approx  - 0.115$, and $\Delta t$ is the characteristic falloff time of the magnitude of the potential with constant rate $\Phi ' \approx {Z_0}$. After this moment in time, winding of the phase trajectory around the zero attractive focus/center begins.  The number of turns of the spiral here is infinite.}
  \label{img:1}
\end{minipage}
\hspace{.02\textwidth}
\begin{minipage}[t]{.48\textwidth}
  \centering
  \includegraphics[width=\textwidth]{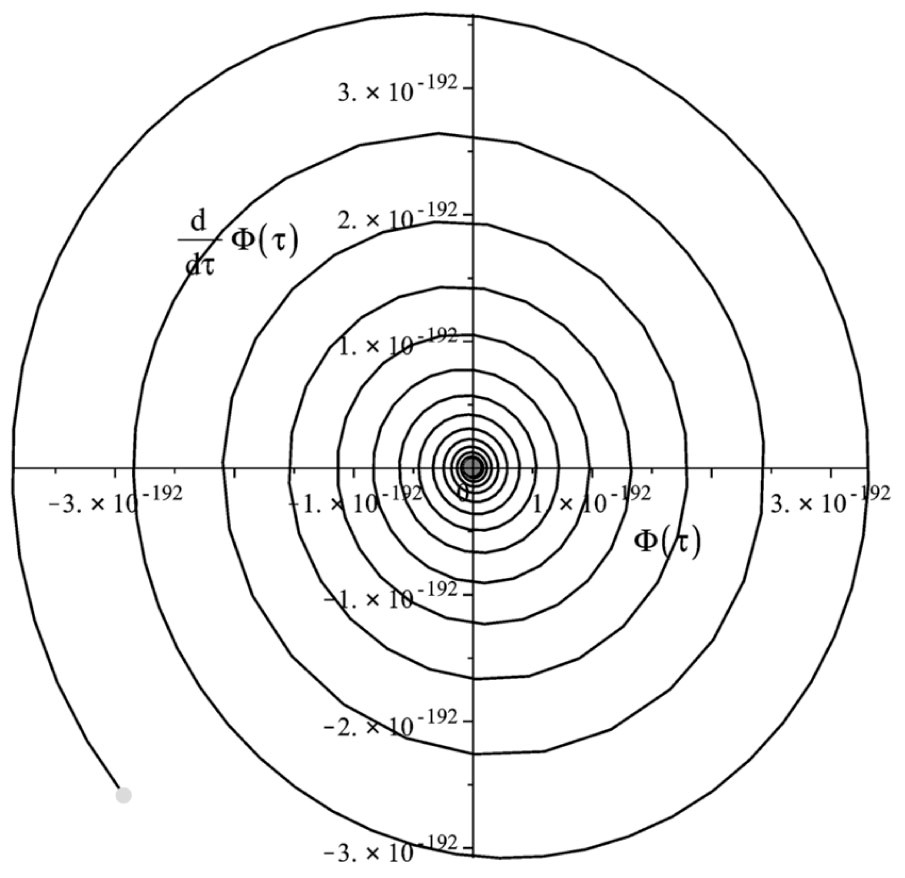}
  \caption{Phase portrait of the system  [1, Eq. (33)] in its final stage in small scale for a classical scalar field in the case ${\alpha _m} = 0$, ${\lambda _m} = 0.001$, $\tau  = 9000 - 10000$, $\Phi (0) = 1$, and $Z(0) = 0$.  The graph was obtained with the help of an enhanced accuracy Runge–Kutta integration method, good to 7–8 orders.}
  \label{img:2}
\end{minipage}
\end{figure}

1c) The case of a large value of the cosmological constant     
\begin{equation}
{\lambda _m} > \frac{4}{3}.
\end{equation}

Here we have two different real and, according to formulas (8), negative eigenvalues  , ${k_1} < 0,{k_2} < 0$.  In this case, the singular point is a stable attractive node. As $\tau  \to  + \infty$ all of the phase trajectories of the dynamical system arrive at the singular point, and all of the phase trajectories, except for just two, as they arrive at the singular point are tangent to the eigenvector ${{\mathbf{u}}_1}$ corresponding to the eigenvector that has the smallest magnitude, i.e., ${k_1}$. Just two trajectories are tangent to the second eigenvector ${{\mathbf{u}}_2}$.  The indicated eigenvectors are equal to 
\begin{equation}
{{\mathbf{u}}_1} = (1,{k_1});\quad {{\mathbf{u}}_1} = (1,{k_2}).
\end{equation}

The angle $\phi$ between the eigenvectors is determined by the relation 
\begin{equation}
\cos \phi  \equiv \frac{{{{\mathbf{u}}_1}{{\mathbf{u}}_2}}}{{\sqrt {{\mathbf{u}}_1^2{\mathbf{u}}_2^2} }} = \sqrt {\frac{4}{{3{\lambda _m}}}}  < 1 \Rightarrow \phi  > 0.
\end{equation}

For very large values of ${\lambda _m}$ the angle between the eigenvectors tends to $\pi /2$ while for ${\lambda _m} \to 4/3$ it tends to zero.  

1d) The degenerate case 
\begin{equation}
{\lambda _m} = \frac{4}{3}.
\end{equation}

This case practically coincides with the previous case, allowing only for the circumstance that all of the trajectories arrive at the singular point tangent to one eigenvector – this in fact corresponds to the above-indicated limit case $\phi  \to 0$. 

\subsubsection{Classical field with repulsion (${e_1} = 1; {e_2} =  - 1$) or phantom field with attraction (${e_1} =  - 1; {e_2} = 1$)}

In this case, regardless of the character of the field (classical / phantom), according to formulas (8) we have two real values with opposite signs.  Thus, the zero singular point (Eqs. (4)) in this case is a saddle point.

\subsection{Characteristic equation and qualitative analysis in the case near a zero singular point  }

Let us calculate the derivatives of the functions described by Eqs. (36) in [1] at the nonzero singular points ${M_ \pm }({x_ \pm },0)$ (Eqs. (5)) for  ${\lambda _m} \geqslant 0$: 
\begin{equation*}
\begin{array}{ll}
  {\displaystyle{{\left. {\frac{{\partial P}}{{\partial x}}} \right|}_{{M_ \pm }}} = 0,}&{\displaystyle{{\left. {\frac{{\partial P}}{{\partial y}}} \right|}_{{M_ \pm }}} = 1,} \\ 
  {\displaystyle{{\left. {\frac{{\partial Q}}{{\partial x}}} \right|}_{{M_ \pm }}} = 2{e_1}{e_2},}&{\displaystyle{{\left. {\frac{{\partial Q}}{{\partial y}}} \right|}_{{M_ \pm }}} =  - \sqrt 3 \sqrt {{\lambda _m} + \frac{1}{{2{\alpha _m}}}} .} 
\end{array}
\end{equation*}

Note that as a consequence of condition (6), the values of the derivatives are real.  The determinant of the matrix of the dynamical system is equal to 
\begin{equation}
{\Delta _ \pm } = {\text{det}}{(A)_{{M_ \pm }}} =  - 2{e_1}{e_2}.
\end{equation}

Thus we obtain the roots of the characteristic equation ${k_ \pm }$, which coincide for the symmetric points ${M_ \pm }$: 
\begin{equation}
\begin{gathered}
  {k_ \pm } = \frac{{\sqrt 3 }}{2}\left[ { - \sqrt {{\lambda _m} + \frac{1}{{2{\alpha _m}}}}  \pm \sqrt {{\lambda _m} + \frac{1}{{2{\alpha _m}}} + \frac{{8{e_1}{e_2}}}{3}} } \right] \hfill \\ 
\end{gathered},
\end{equation}
such that    
\begin{equation}
{k_1}{k_2} =  - 2{e_1}{e_2}.
\end{equation}

\subsubsection{Classical field with attraction (${e_1} = {e_2} = 1$) or phantom field with repulsion  (${e_1} = {e_2} =  - 1$) }

As a consequence of condition (6), the radicand (expression under the radical) in the first term of expression (18) is strictly greater than zero; moreover, as a consequence of the fact that ${e_1}{e_2} =  + 1$ the radicand in the second term of expression (18) is greater than the radicand in the first term; therefore, both eigenvalues are real and opposite in sign. Thus, the points ${M_ \pm }$ in this case are unstable saddle points for either direction of time. 

\subsubsection{Classical field with repulsion (${e_1} = 1,\,\,{e_2} =  - 1$) or phantom field with attraction  (${e_1} =  - 1,\,\,{e_2} = 1$)}

As a consequence of condition (6), the radicand (expression under the radical) in the first term of expression (18) is strictly greater than zero; moreover, as a consequence of the fact that ${e_1}{e_2} =  - 1$ the radicand in the second term of expression (18) is less than the radicand in the first term and, generally speaking, can also be negative.  Therefore, three cases are possible: 

1) ${\lambda _m} + 1/2{\alpha _m} - 8/3 > 0$ -- thus, both eigenvalues are real and negative.  In this case, the solution contains two symmetric attractive (stable) nondegenerate nodes.  All of the phase trajectories in the neighborhoods of such singular points converge on these points as $t \to \infty$ and, with the exception of only two trajectories, are tangent to the shortest eigenvector. 

2) ${\lambda _m} + 1/2{\alpha _m} - 8/3 = 0$ -- thus, both eigenvalues are negative and equal. In this case, the solution contains two symmetric degenerate nodes, which are bifurcation points of the dynamical system.
  
3) ${\lambda _m} + 1/2{\alpha _m} - 8/3 < 0$ -- thus, the two eigenvalues are complex conjugate, and their real parts are negative.  In this case, the solution contains two symmetric attractive foci. 
In the case of two symmetric foci, it is easy to find the limiting value ${H_m}(\infty )$ to which the Hubble constant tends in the limit $t \to \infty$.  Substituting the coordinates of the foci ${M_ \pm }( \pm 1/\sqrt {{e_2}\alpha } ,0)$ into equation [1, Eqs. (32) ], we obtain  
\begin{equation}
{H_m}(\infty ) = \sqrt {\frac{1}{3}\left( {{\lambda _m} - \frac{1}{{2{\alpha _m}}}} \right)}.
\end{equation}

\section{NUMERICAL MODELING OF COSMOLOGICAL EVOLUTION FOR TYPICAL CASES}

Results of numerical modeling in the mathematical computing system Mathematica confirmed the results of qualitative analysis and established the asymptotic behavior of the trajectories at infinity.  Let us consider some examples. 

\subsection{Case of a pair of saddle singular points (${e_1} = {e_2} = 1$, ${e_1} = {e_2} =  - 1$)}

In the real region of the solution (that part of it conforming to condition (6)) in the case of identical values of the parameters ${e_1}$ and ${e_2}$ the system has three singular points: a pair of symmetric saddle points with coordinates ${M_ \pm }\left( { \pm \frac{1}{{\sqrt {{e_2}{\alpha _m}} }},0} \right)$ and the zero singular point \[{M_0}(0,0)\]. Figure 3 displays the topological structure of the phase trajectories with an attractive focus at the origin.  Figure 4 displays phase trajectories with an attractive center at the origin.

\begin{figure}[ht!]
\centering
\begin{minipage}[t]{.48\textwidth}
  \centering
  \includegraphics[width=\textwidth]{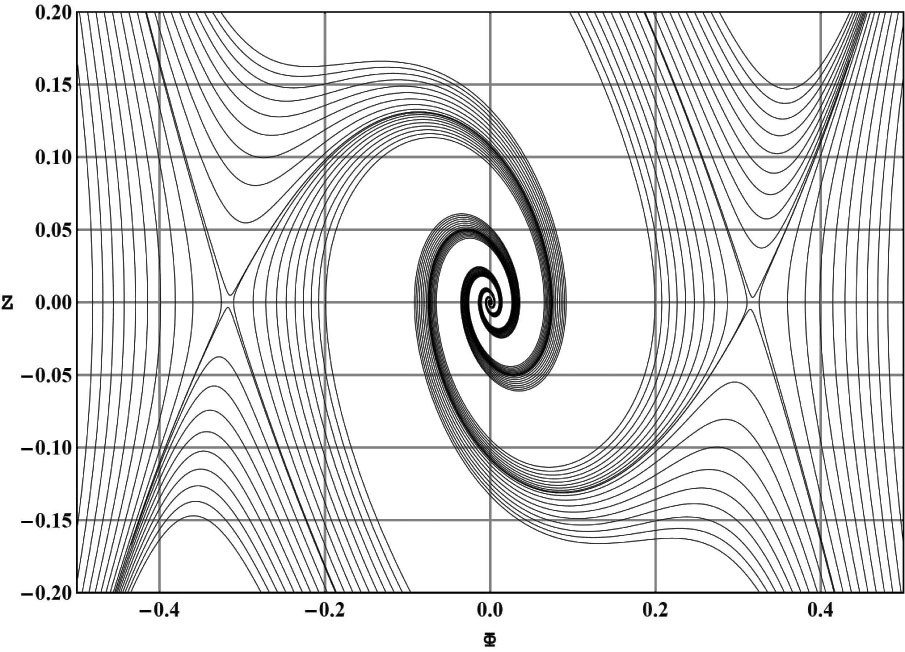}
  \caption{Phase trajectories of the system [1, Eqs. (36)] (${e_1} = 1$, ${e_2} = 1$, ${\alpha _m} = 10$, ${\lambda _m} = 0.1$).  The allowed region is shown in Fig. 3a  of  [1].}
  \label{img:3}
\end{minipage}
\hspace{.02\textwidth}
\begin{minipage}[t]{.48\textwidth}
  \centering
  \includegraphics[width=\textwidth]{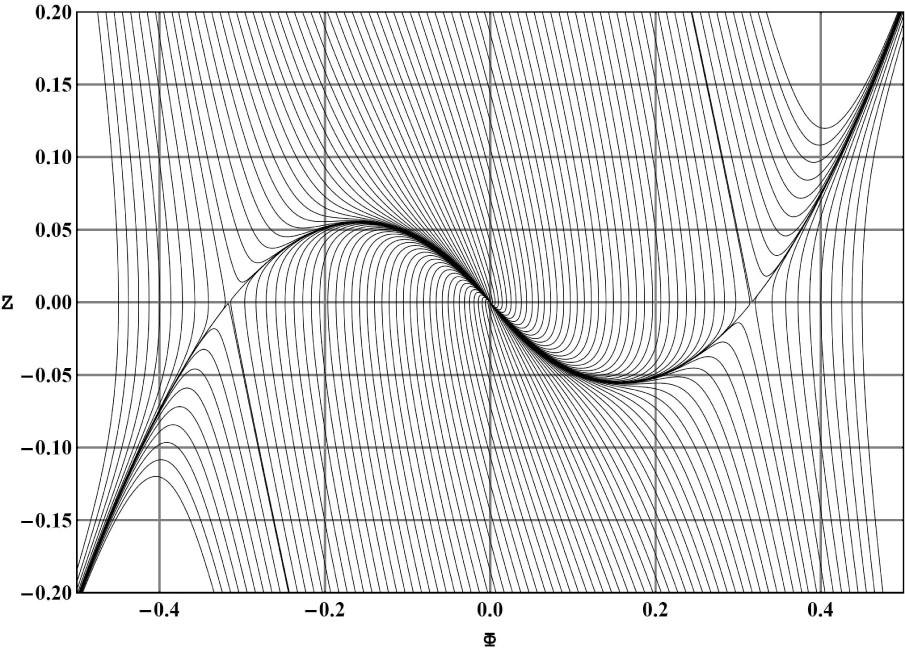}
  \caption{Phase portraits of the trajectory of the system [1, Eqs. (36)] (${e_1} = 1$, ${e_2} = 1$, ${\alpha _m} = 10$, ${\lambda _m} = 1.5$).}
  \label{img:4}
\end{minipage}
\end{figure}

\subsection{Case of a zero saddle singular point  (${e_1} = 1$, ${e_2} =  - 1$, ${e_1} =  - 1$, ${e_2} = 1$)}

If the conditions of realness of the solution for classical and phantom scalar fields are satisfied in the case of different signs of the parameters ${e_1}$ and ${e_2}$, the system  [1, Eqs. (36)] has the following singular points: a zero saddle singular point and a pair of symmetric points whose character is determined by the parameters ${e_1}$, ${\alpha _m}$, and ${\lambda _m}$.  For numerical interpretation of dynamical system [1, Eqs. (38)] it is necessary to choose initial values of the functions $\Phi$ and $Z$ satisfying condition [1, Eqs. (38)].  Figure 5 displays phase trajectories for a classical scalar field.  It is clear that the regions near the symmetric attractive foci do not satisfy condition [1, Eqs. (38)]; therefore, the trajectories having their starting point at $\tau  =  - \infty$ (pole) terminate at the boundary of this region.  Figure 6 presents a phase portrait of the system for a phantom scalar field with the same parameter ${\lambda _m}$.  It can be seen that now regions far from the symmetric attractive foci are inaccessible.  In the given example ${\lambda _m} = 0$, and the allowed region is divided by the directrices of the saddle point into two regions, each of which contains its own family of phase trajectories. 

\begin{figure}[ht!]
\centering
\begin{minipage}[t]{.48\textwidth}
  \centering
  \includegraphics[width=\textwidth]{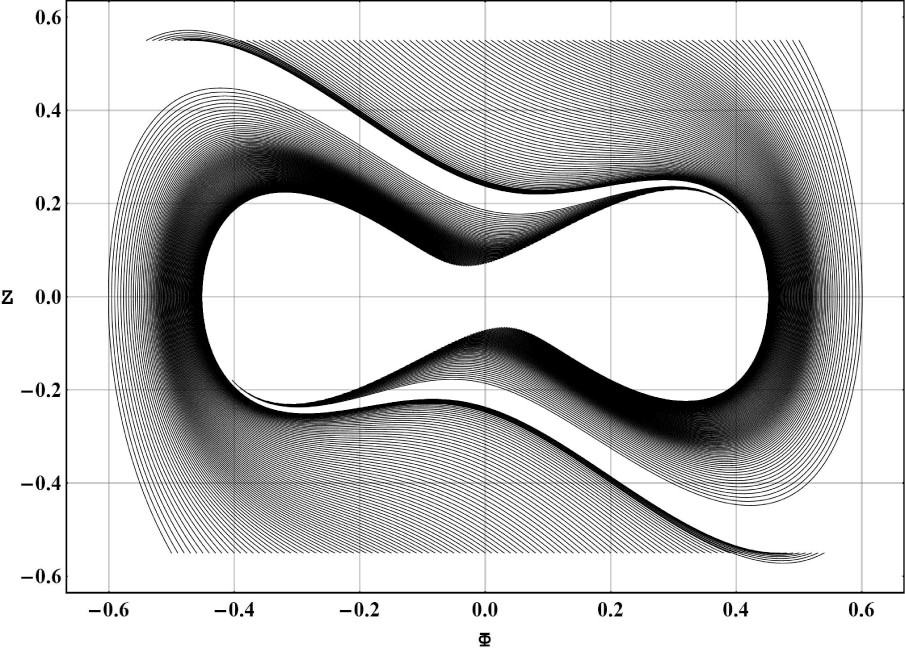}
  \caption{Phase trajectories of system [1, Eqs. (36)] for a classical field: ${e_1} = 1$, ${e_2} =  - 1$, ${\alpha _m} =  - 10$, and ${\lambda _m} = 0$.  The allowed region is shown in Fig. 2a [1].}
  \label{img:5}
\end{minipage}
\hspace{.02\textwidth}
\begin{minipage}[t]{.48\textwidth}
  \centering
  \includegraphics[width=\textwidth]{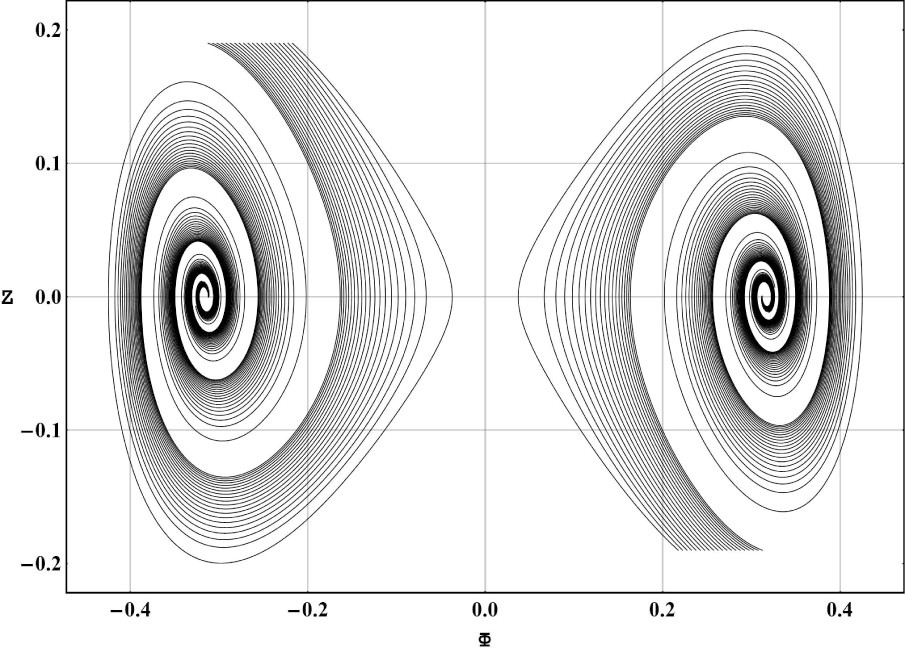}
  \caption{Phase trajectories of system [1, Eqs. (36)] for a phantom field: ${e_1} =  - 1$, ${e_2} = 1$, ${\alpha _m} = 10$, and ${\lambda _m} = 0$.}
  \label{img:6}
\end{minipage}
\end{figure}

\begin{figure}[ht!]
\centering
\begin{minipage}[t]{.48\textwidth}
  \centering
  \includegraphics[width=\textwidth]{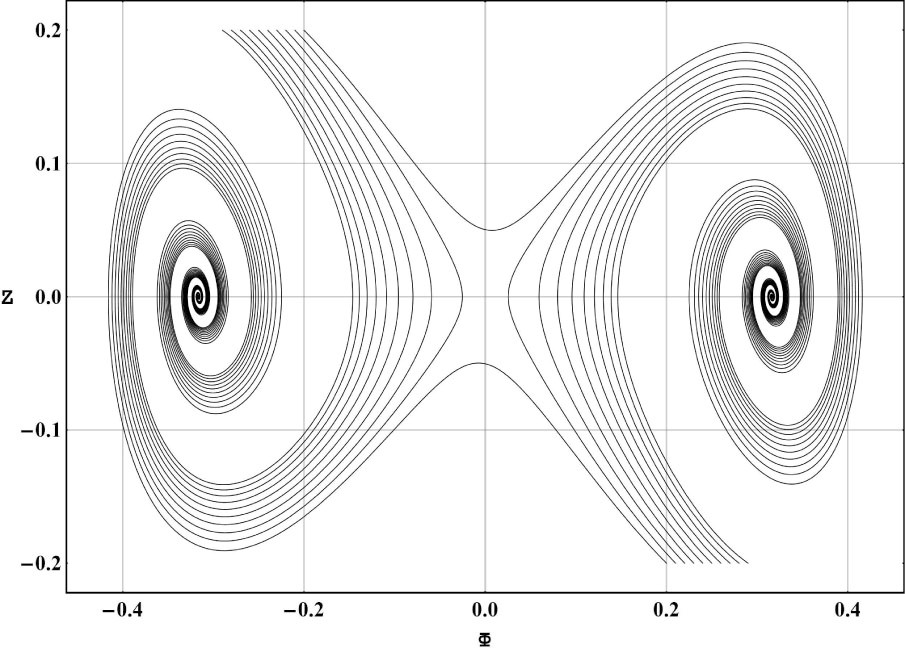}
  \caption{Phase trajectories of system [1, Eqs. (36)] for a phantom field: ${e_1} =  - 1$, ${e_2} = 1$, ${\alpha _m} = 10$, and ${\lambda _m} = 0.01$.}
  \label{img:7}
\end{minipage}
\hspace{.02\textwidth}
\begin{minipage}[t]{.48\textwidth}
  \centering
  \includegraphics[width=\textwidth]{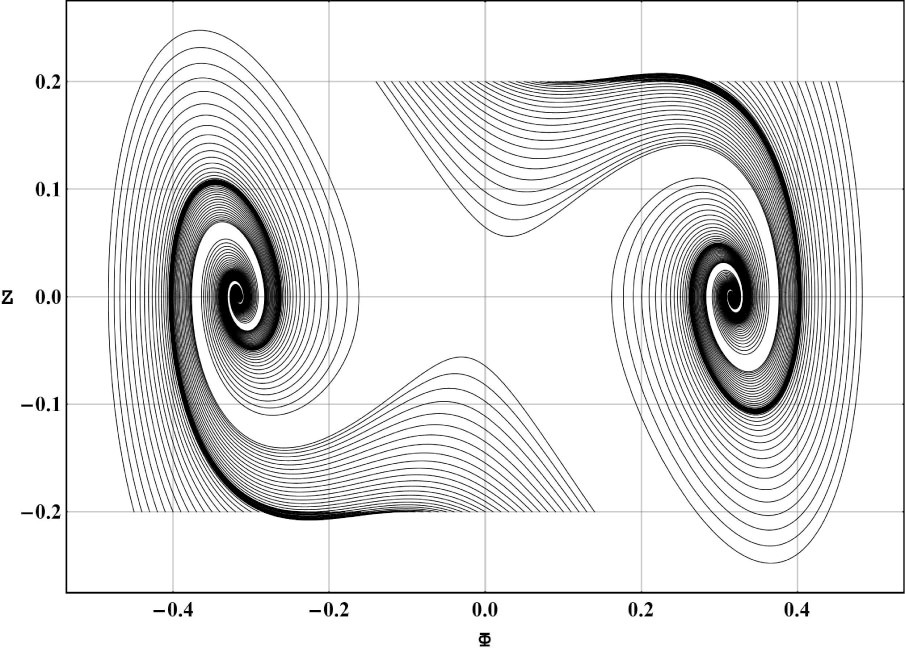}
  \caption{Phase trajectories of system [1, Eqs. (36)] for a phantom field: ${e_1} =  - 1$, ${e_2} = 1$, ${\alpha _m} = 10$, and ${\lambda _m} = 0.1$.}
  \label{img:8}
\end{minipage}
\end{figure}

For nonzero values of the parameter ${\lambda _m}$ the allowed region contains phase trajectories, some of which pass out of the neighborhood of one focus into the neighborhood of the other (Figs. 7 and 8). With further increase of the parameter ${\lambda _m}$ the trajectories near the symmetric foci become distorted (Fig. 9).  Figure 10 shows the case of bifurcation of a dynamical system upon changeover of the character of the symmetric singular points from an attractive focus to an attractive center corresponding to the value ${\lambda _m} = 2.7167$. 

\begin{figure}[ht!]
\centering
\begin{minipage}[t]{.48\textwidth}
  \centering
  \includegraphics[width=\textwidth]{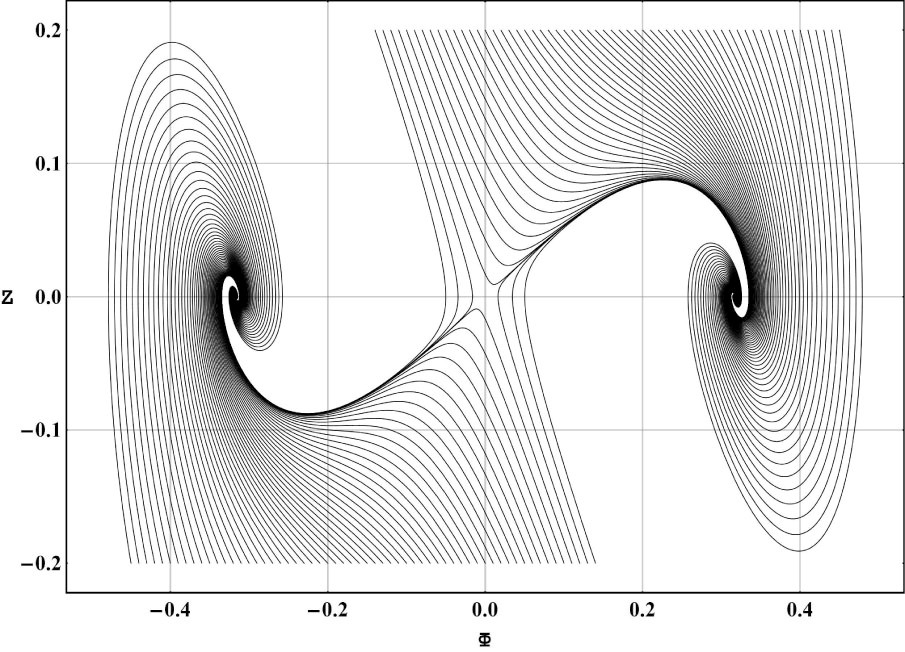}
  \caption{Phase trajectories of system [1, Eqs. (36)] for a phantom field: ${e_1} =  - 1$, ${e_2} = 1$, ${\alpha _m} = 10$, and ${\lambda _m} = 0.5$.}
  \label{img:9}
\end{minipage}
\hspace{.02\textwidth}
\begin{minipage}[t]{.48\textwidth}
  \centering
  \includegraphics[width=\textwidth]{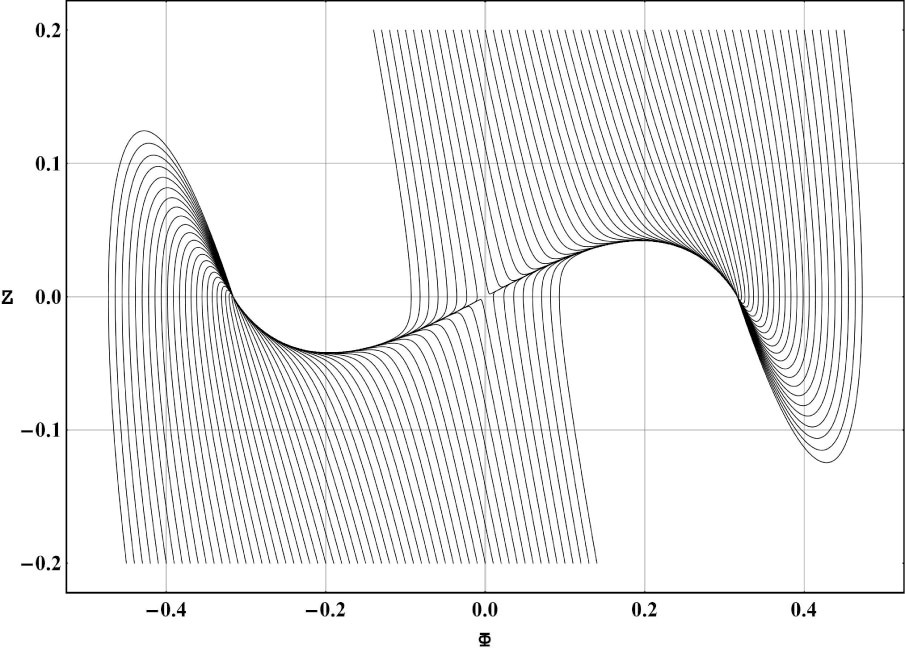}
  \caption{Phase trajectories of system [1, Eqs. (36)] for a phantom field: ${e_1} =  - 1$, ${e_2} = 1$, ${\alpha _m} = 10$, and ${\lambda _m} = 2.7167$.  Bifurcation.}
  \label{img:10}
\end{minipage}
\end{figure}

\section{DISCUSSION OF RESULTS}

We note to start with that the coordinates in the Poincar\'e diagrams (Figs. 11-14) of $\Phi$ and $Z$ are related to the dynamical variables $x$ and $y$ as follows:   
\begin{equation*}
\Phi  = \frac{x}{{\sqrt {1 + {x^2} + {y^2}} }},\,\,\,\,\;Z = \frac{y}{{1 + \sqrt {{x^2} + {y^2}} }};
\end{equation*}
therefore, infinitely large values of the dynamical variables $\Phi  \to  \pm \infty$, $\dot \Phi  \to  \pm \infty$ correspond to points on the unit circle.  Each specific phase trajectory is determined by some point on the unit circle (initial conditions), and the direction of motion along a phase trajectory is determined by the direction from the corresponding pole to this point.

\begin{figure}[ht!]
\centering
\begin{minipage}[t]{.48\textwidth}
  \centering
  \includegraphics[width=\textwidth]{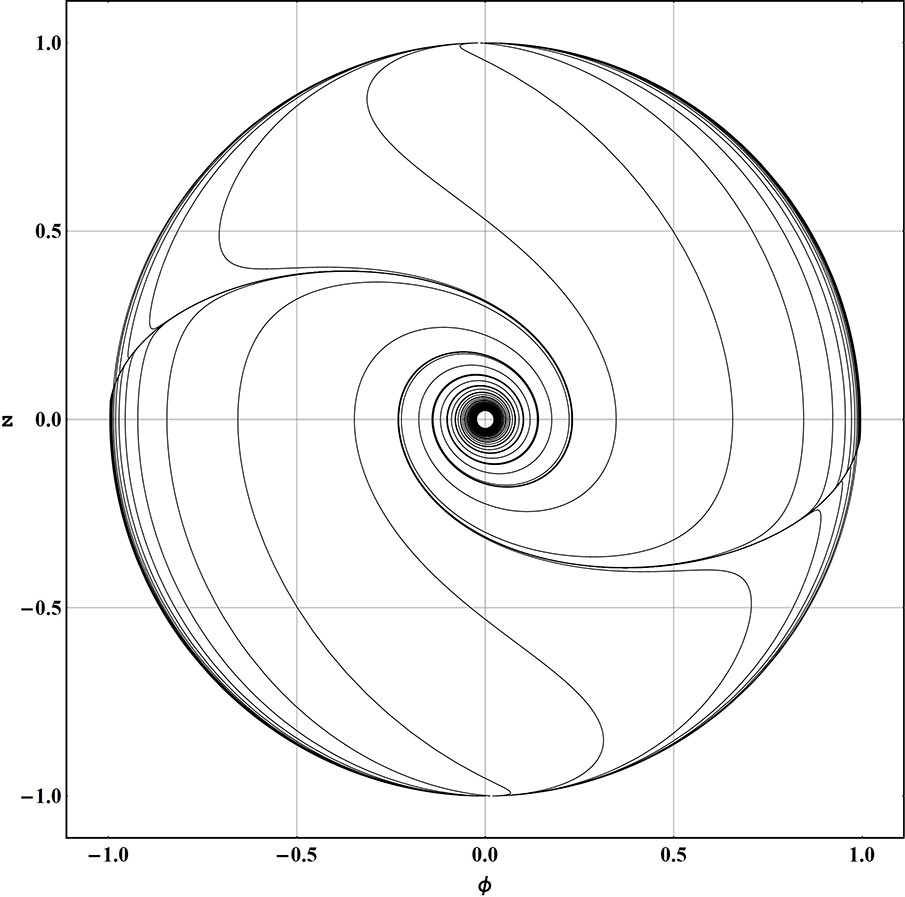}
  \caption{Phase trajectories of system [1, Eqs. (36)] on the Poincar\'e sphere for a classical field without self-action for ${e_1} = 1$, ${e_2} = 1$, ${\alpha _m} = 0$, and ${\lambda _m} = 0$.}
  \label{img:11}
\end{minipage}
\hspace{.02\textwidth}
\begin{minipage}[t]{.48\textwidth}
  \centering
  \includegraphics[width=\textwidth]{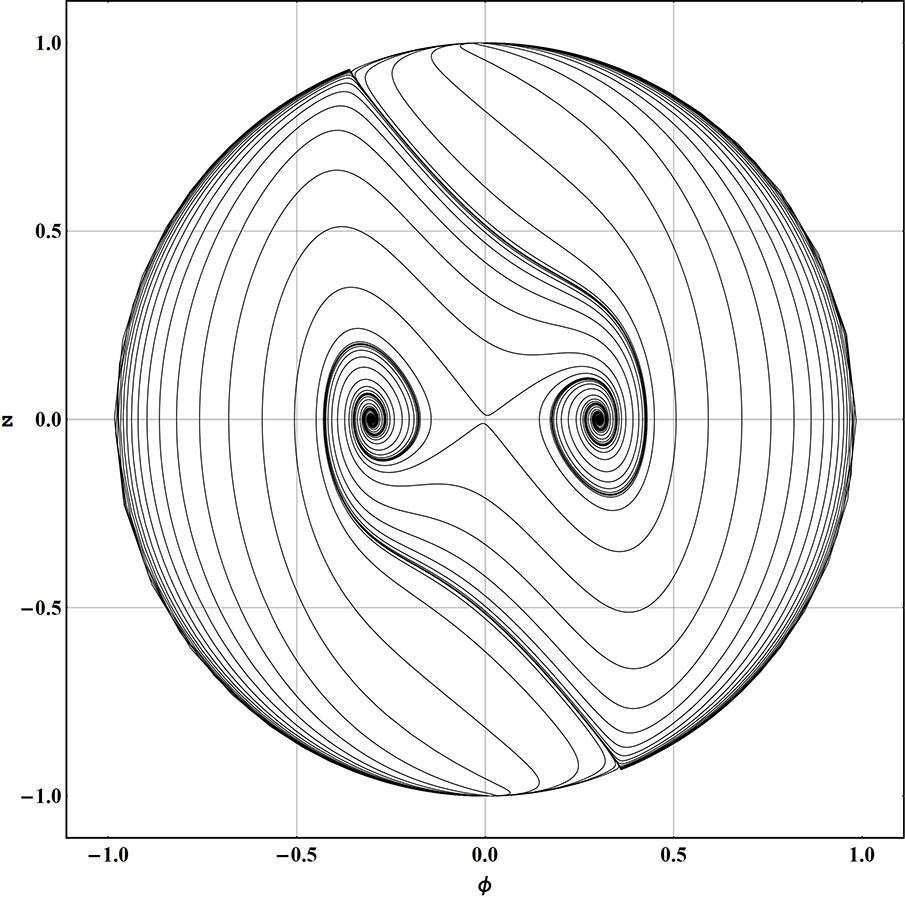}
  \caption{Phase trajectories of system [1, Eqs. (36)] on the Poincar\'e sphere for a classical field with self-action for ${e_1} = 1$, ${e_2} =  - 1$, ${\alpha _m} =  - 10$, and ${\lambda _m} = 0.1$.}
  \label{img:12}
\end{minipage}
\end{figure}

\begin{figure}[ht!]
\centering
\includegraphics[width=.6\textwidth]{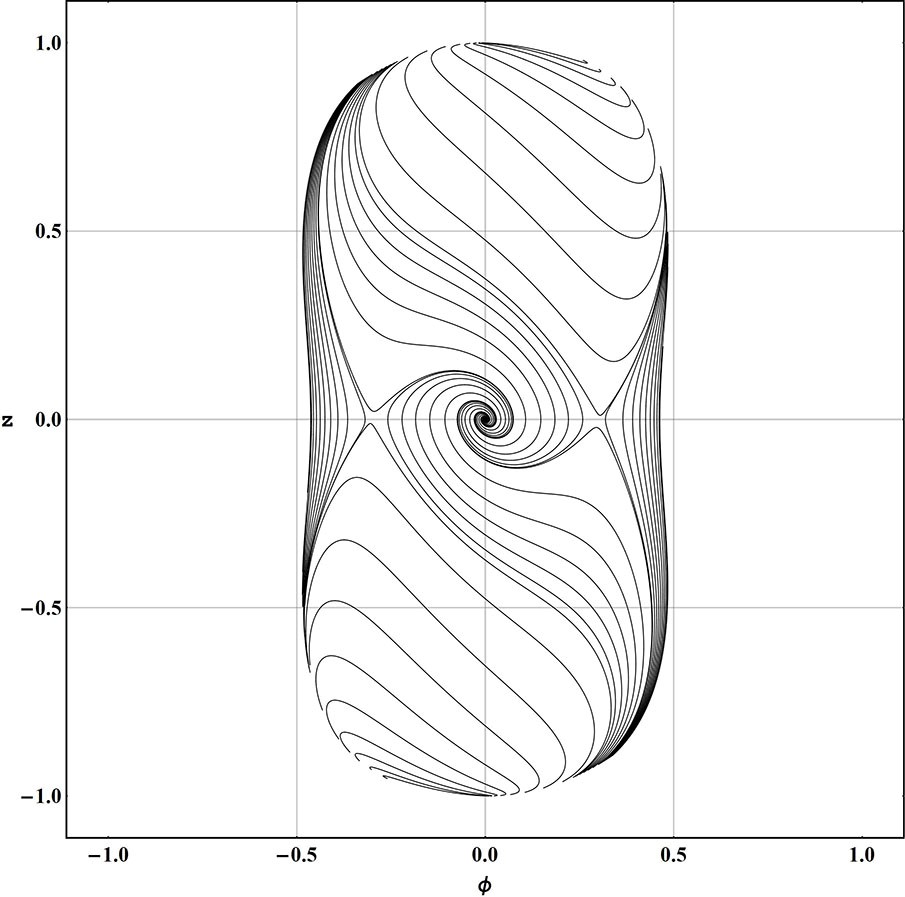}
\caption{Phase trajectories of system [1, Eqs. (36)] on the Poincar\'e sphere for a classical field with self-action for ${e_1} = 1$, ${e_2} = 1$, ${\alpha _m} = 10$,  and ${\lambda _m} = 0.1$.}
\label{img:13}
\end{figure}

\begin{figure}[ht!]
\centering
\includegraphics[width=.95\textwidth]{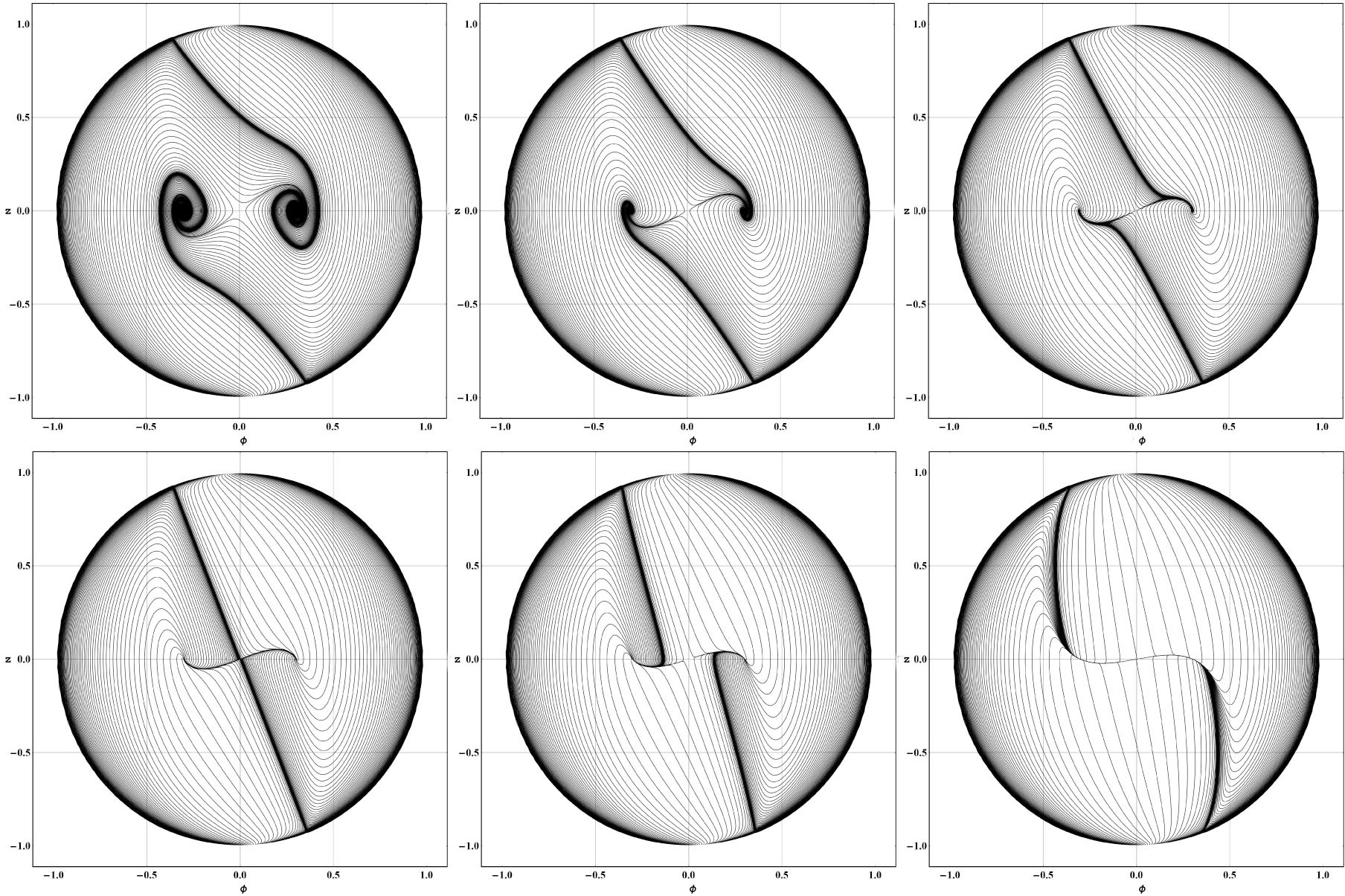}
\caption{Phase trajectories of system [1, Eqs. (36)] of a classical scalar field on the Poincar\'e sphere for the parameters ${e_1} = 1$, ${e_2} =  - 1$, and ${\alpha _m} =  - 10$; upper row, from left to right: ${\lambda _m} = 0.1,\;0.5,\;1.0$; lower row, from left to right: ${\lambda _m} = 1.61666,\;2.61666,\;10$.}
\label{img:14}
\end{figure}

Comparing the projections of the phase portraits of the dynamical systems based on a classical scalar field with self-action (Figs. 11–14), we note the following:  

– All of the phase portraits possess central symmetry $\{ \Phi ,Z\}  \to \{  - \Phi , - Z\}$.  This allows us to analyze the phase trajectories starting out in one half-plane, say for definiteness, the upper half-plane. 

– In the case of zero value of the cosmological constant and absence of self-action ($\alpha  = 0$) all of the phase trajectories are separated by separatrices emanating from the equator $\{ {\Phi _\infty } =  \pm \infty ,\,\,\,{Z_\infty } = 0\}$ into two dynamical flows emanating from the poles $\{ {\Phi _\infty } = 0,\,\,\,{Z_\infty } =  \pm \infty \}$ and asymptotically wrapping themselves around the focus $\{ 0,0\}$ as $t \to  + \infty$ (Fig. 11). 

– In the case $\lambda  > 0$, ${e_2} = 1$, and $\alpha  > 0$ one attractive center and two symmetric saddle points arise: all of phase space is partitioned into six dynamical flows, and a pair of them emanate from the poles and are incident upon the opposite poles. The other pair, emanating from the poles, turns in the opposite direction, and finally, the third pair of flows, emanating from the poles, winds around the attractive center (Fig. 13). 

–  Finally, there exists for $\lambda  > 0$, ${e_2} =  - 1$, and $\alpha  < 0$ one more transitional type of phase diagram – a bifurcation diagram, where all trajectories divide into two dynamical flows emanating from the poles and turning around after a quite complex loop (Fig. 14). 

– In the case $\lambda  > 0$, ${e_2} =  - 1$, and $\alpha  < 0$ two attractive centers and one saddle central point arise: all of phase space is partitioned into four dynamical flows emanating from the poles, where the flow emanating from each pole divides into a pair of flows, the trajectory of one of which winds around the left attractive center, and that of the other one, around the right attractive center (Fig. 12).  

Let us turn our attention to the most interesting case of the model $\lambda  \geqslant 0$ with a classical scalar field and self-action ${e_1} = 1$, ${e_2} =  - 1$, and $\alpha  < 0$,  the phase trajectories of which are depicted in Fig. 5.  As can be seen from this figure, the phase trajectories emanating from infinity over the course of time approach the boundary of the region of accessible values of the dynamical variables.  It needs to be understood what happens with these trajectories.  To start with, we observe that the scalar field equation [1, Eqs. (6)] for a Friedmann Universe has the form of the standard equation of nonlinear oscillations with a friction coefficient $\beta (x,x')$ that depends on the coordinate and velocity of the particle taken into account (see, for example, [5]):
\begin{equation}
x'' + \beta x' + V'(x) = 0,
\end{equation} 
where  $\beta = 3 H_m (x) = \sqrt{3\varepsilon_m}$.  Therefore, the approach of this trajectory to the line ${\varepsilon_{{\text{eff}}}} = 0$ can be interpreted as a falloff to zero of the friction coefficient near the inaccessible region.  But then the trajectory of the system must transition into a trajectory of free nonlinear potential oscillations: 
\begin{equation}
x'' + V'(x) = 0,
\end{equation} 
having a total energy integral 
\begin{equation}
\frac{{{{x'}^2}}}{2} + V(x) = {E_0}.
\end{equation} 

It is namely this integral that describes the curve [1, Eqs. (39)] if we renormalize the constant   with the help of the cosmological constant.  Figure 15 displays the results of numerical integration of the equation of free oscillations [1, Eqs. (68)] while Fig. 16 displays the results of numerical integration of the dynamical equations [1, Eqs. (36)] near a trajectory of free oscillations.  Thus, over a quite wide region of the parameters of the model of a nonlinear scalar field, classical as well as phantom, the emergence of a limit cycle in the limit   is possible, described by a trajectory with zero effective energy [1, Eqs. (39)] and representing free (nondecaying) nonlinear oscillations in a potential field.    Naturally, the question arises as to the energy and pressure of the scalar field in this final state. Regardless of the parameters of the model, substituting an expression for the energy density of the scalar field [1, Eqs. (15)] into the equation of the limit trajectory [1, Eqs. (39)], we obtain for the energy density of the scalar field on the limit curve the value 
\begin{equation}
{\varepsilon _\infty } =  - \frac{\lambda }{{8\pi }}.
\end{equation} 

Thus, for   this quantity should be negative, which is possible only for certain values (and signs!) of the model parameters. Substituting the found value of the energy density into expression [1, Eq. (15)] for the pressure of the scalar field, we find on the limit curve the value 
\begin{equation}
{p_\infty } = \frac{1}{{8\pi }}(2{m^2}{e_1}{\Phi '^2} + \lambda ).
\end{equation}

\begin{figure}[ht!]
\centering
\begin{minipage}[t]{.48\textwidth}
  \centering
  \includegraphics[width=\textwidth]{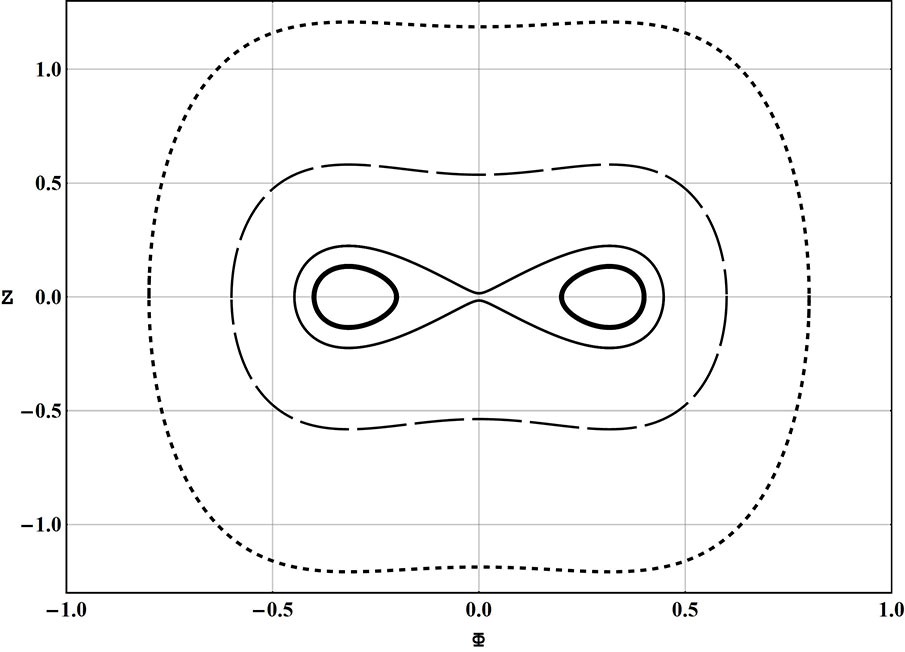}
  \caption{Phase trajectories of free oscillations [1, Eqs. (68)]: ${e_1} = 1$, ${e_2} =  - 1$, $\alpha  =  - 10$, the thick curve corresponds to $\lambda  = 0.288$, the thin curve corresponds to $\lambda  = 0.400769$, the dashed curve, to $\lambda  = 1.008$, and the dotted curve, to $\lambda  = 2.688$.}
  \label{img:15}
\end{minipage}
\hspace{.02\textwidth}
\begin{minipage}[t]{.48\textwidth}
  \centering
  \includegraphics[width=\textwidth]{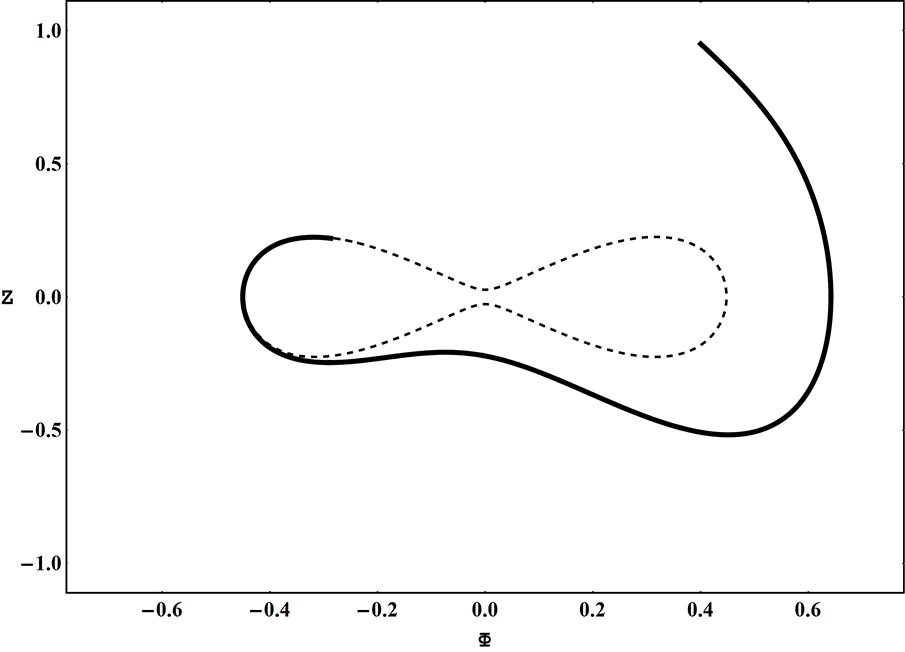}
  \caption{Approach of the phase trajectory of dynamical system [1, Eq. (36)] in the case of a classical scalar field (thick, continuous curve) to the trajectory of free potential oscillations (Eq. (68)) (dotted curve): ${e_1} = 1$, ${e_2} =  - 1$, $\alpha  =  - 10$ and $\lambda  = 0$.  The allowed region is shown in Fig. 2a [1].}
  \label{img:16}
\end{minipage}
\end{figure}

Let us now investigate the asymptotic behavior of the system in the case under consideration as $\tau  \to  + \infty$.  Since $H = \dot aa \to 0$, it follows that $\dot a \to 0$; consequently, it also follows that $\ddot a \to 0$.  But then as a consequence of the Einstein equation for the spatial components the following limits should hold:  
\begin{equation*}
p \to \frac{\lambda }{{8\pi }},\;t \to  + \infty.
\end{equation*}

According to [1, Eqs. (15) and (27)], this is possible only if the conditions $\dot \Phi  \to 0 \to \Phi  \to {\Phi _0}$ and ${\varepsilon _{{\text{eff}}}} = 0$ are met:
\begin{equation*}
{e_2}{m^2}\Phi _0^2 - \frac{\alpha }{2}\Phi _0^4 + \lambda  = 0 \Rightarrow {e_2}x_0^2 - \frac{{{\alpha _m}}}{2}x_0^4 + {\lambda _m} = 0.
\end{equation*}
It is easy to see that this equation coincides with the equation of the curve [1, Eqs. (39)] for $y = 0$; consequently, it determines a point lying at the intersection of the abscissa with this curve, and can be rewritten in terms of the potential  $V(\Phi )$ [1, Eqs. (2)]: 
\begin{equation}
V({\Phi _0}) =  - \frac{1}{2}\left( {\frac{1}{{2\alpha }} + \lambda } \right).
\end{equation}

As can be seen from [1, Fig. 1], depending on the parameters of the model, there can be two such points: ${M_0}( \pm {\Phi _0},0)$ or four.  Thus, in the limit $t \to  + \infty$ for the appropriate parameters of the model, such a Universe can tend to a Euclidean Universe with a scalar vacuum completely compensating the cosmological term.  The process of transition to this expansion regime requires further study.

The work was performed in accordance with the Russian Government Program of Competitive Growth of Kazan Federal University.


\end{otherlanguage}

%

\end{document}